\documentclass[5p]{elsarticle}

\usepackage{times}
\usepackage{graphicx}
\usepackage{url}
\usepackage{booktabs, makecell, tabularx}
\usepackage{threeparttable}
\usepackage{array,multirow}
\usepackage{textcomp}
\usepackage{enumitem}
\usepackage[table]{xcolor}
\usepackage[hidelinks, bookmarks=false]{hyperref}

\journal{%
~\copyright~2023. This manuscript version is made available under the CC-BY-NC-ND 4.0 license 
\href{https://creativecommons.org/licenses/by-nc-nd/4.0/}{\color{blue}{https://creativecommons.org/licenses/by-nc-nd/4.0/}}}

\begin{document}
\begin{frontmatter}


\title{Accelerating AI and Computer Vision for\\Satellite Pose Estimation on the Intel Myriad X Embedded SoC}

\author[1]{Vasileios~Leon}
\cortext[tgtgt]{Accepted for publication at \emph{Elsevier Microprocessors and Microsystems}, Vol. 103, Nov. 2023, DOI:
\href{https://doi.org/10.1016/j.micpro.2023.104947}{\color{blue}{https://doi.org/10.1016/j.micpro.2023.104947}}}
\author[1]{Panagiotis~Minaidis}
\author[2,1]{George~Lentaris}
\author[1]{Dimitrios~Soudris}

\affiliation[1]{organization={National Technical University of Athens, School of Electrical and Computer Engineering},
country={Greece}}

\affiliation[2]{organization={University of West Attica, Department of Informatics and Computer Engineering},
country={Greece}}


\begin{abstract}
The challenging deployment of Artificial Intelligence (AI) and Computer Vision (CV) algorithms at the edge pushes the community of embedded computing to examine heterogeneous System-on-Chips (SoCs). Such novel computing platforms provide increased diversity in interfaces, processors and storage, however, the efficient partitioning and mapping of AI/CV workloads still remains an open issue. In this context, the current paper develops a hybrid AI/CV 
system on Intel's Movidius Myriad X, which is an heterogeneous Vision Processing Unit (VPU), for initializing and tracking the satellite's pose in space missions. The space industry is among the communities examining alternative computing platforms to comply with the tight constraints of on-board data processing, while it is also striving to adopt functionalities from the AI domain. At algorithmic level, we rely on the ResNet-50-based UrsoNet network along with a custom classical CV pipeline. For efficient acceleration, we exploit the SoC's neural compute engine and 16 vector processors by combining multiple parallelization and low-level optimization techniques. The proposed single-chip, robust-estimation, and real-time solution delivers a throughput of up to 5 FPS for 1-MegaPixel RGB images within a limited power envelope of 2W.
\end{abstract}


\end{frontmatter}

\section{Introduction}
The rapid evolution of
powerful Artificial Intelligence (AI) algorithms
and sophisticated Computer Vision (CV) pipelines
marks a new era of computing at the edge. 
The worldwide demand for high performance 
with a restricted power budget,
especially when considering embedded systems,
challenges the integration of 
AI/CV functionalities in novel applications.
Heterogeneous System-on-Chip (SoC) processors 
emerge as a promising solution \cite{heterog}, 
providing increased programming flexibility
and diversity in terms of processors and memories.
A prominent class of heterogeneous SoCs
is the Vision Processing Unit (VPU) 
\cite{myriad_vpu, myriad_hotc, paralos},
which excels in embedded low-power imaging applications,
covering domains such as robotics, automotive, and space.

Space is one of the communities
seeking alternative processing platforms 
to comply with the tight constraints of on-board processing.
Towards this direction,
the enhanced performance of modern edge devices 
can efficiently serve tasks for 
Earth Observation (EO) and Vision-Based Navigation (VBN).
Satellite Pose Estimation \cite{ursonet, lourakis, pose1, pose2}
constitutes a typical problem in such space applications,
with its increased computing intensity stressing 
the conventional processors
to provide
sufficient throughput. 
Moreover,
the heterogeneity of devices such as the VPUs,
allows for 
improved adaptability to various mission scenarios
and seamless in-flight re-programmability.
To further improve 
the Performance and Size-Weight-Power (SWaP),
as well as additional 
costs (e.g., development effort),
the space industry is studying 
mixed-criticality architectures \cite{architectures, criticality2, criticality1, asip_2022},
i.e., the integration of both 
space-grade and Commercial-Off-The-Shelf (COTS) components.
The use of COTS components 
in Low Earth Orbit (LEO) missions and CubeSats 
relies on 
the partial shielding provided by the Earth’s magnetosphere
and/or the short mission lifetime,
which limit the damage or unavailability of electronics
due to radiation. 
In this context,
FPGAs \cite{mpsoc, lentarisTVID, iturbe, csp}, 
GPUs \cite{gpu, gpu2, gpu_fpga, gpu_fpga_vpu, kosmidis_2022} and 
VPUs \cite{myriad2, fsat_myriad, hpcb1, hpcb}
are evaluated as accelerators,
while COTS devices, e.g., Intel's Myriad 2 VPU,
are subjected to radiation tests \cite{ai_space}.
A second challenge for the space industry 
is the wider adoption of AI,
which is currently limited to 
offline/ground data processing
instead of on-board processing,
mostly due to insufficient computational power
and qualification issues when deployed in orbit
\cite{ai_space}.

In this paper,
motivated by the aforementioned challenges 
in embedded computing and space avionics, 
we develop AI and CV functionalities
for satellite pose estimation
on Intel's 16-nm Myriad X VPU.
Myriad X is the latest variant of Intel's VPU family,
supporting custom and automatic AI deployment,
as well as classic CV workloads.
In particular,
we accelerate 
a Deep Neural Network (DNN) of ResNet backbone,
namely UrsoNet \cite{ursonet},
which estimates the satellite's pose,
next to
a custom CV pipeline \cite{lourakis}
tracking the pose. 
We select UrsoNet \cite{ursonet} and the CV of \cite{lourakis}
because they are representative of high-performance state-of-the-art algorithms for the pose estimation/tracking problem. 
The two algorithmic pipelines are acting complementary
to provide a single-chip solution 
to the entire pose estimation problem 
(initialization and tracking)
and increase the robustness of the VBN subsystem. 
The experimental results show that the proposed
low-level optimization and tuning
improves the speedup 
provided by the default core parallelization.
Overall,  
we perform pose estimation \& tracking 
on 1-MegaPixel RGB images
with a throughput of up to 5 FPS,
while the power consumption of the SoC is around 2W. 
Moreover,
we directly compare our implementations on the Myriad X VPU
to other competitive embedded devices
(CPUs, embedded GPUs, ARM-based FPGAs).
The contribution of our work lies in two directions:

\begin{itemize}
    \item[(i)] \emph{Embedded Computing}: 
    We propose a certain hybrid embedded system 
    combining both AI and classic CV functions, while 
    we accelerate them
    on an heterogeneous SoC 
    by applying various low-level optimization techniques.
    \item[(ii)] \emph{Space Avionics}: 
    We propose a self-contained system 
    for the entire relative pose estimation problem,
    i.e., Lost-In-Space and Tracking.
    Our solution is integrated in a single COTS SoC 
    and provides excellent power consumption and sufficient performance. 
    We evaluate a candidate device 
    for on-board acceleration  
    in mixed-criticality avionics architectures, 
    while we examine practical AI functionalities for on-board processing 
    (not widely adopted yet). 
\end{itemize}

The remainder of this paper is organized as follows. 
Section \ref{cots} provides background in the use of COTS devices in space,
while Section \ref{vpu}
introduces the Myriad VPU family.
Section \ref{prop} introduces the proposed embedded AI/CV system, and it also 
discusses its mapping onto the Myriad X SoC and low-level implementation details.
Section \ref{exp} reports and analyzes the experimental results,
and finally, 
Section \ref{concl} draws the conclusions. 


\section{COTS Embedded Processors in Space}
\label{cots}
The literature includes a 
variety of COTS-based processing architectures 
and works evaluating the suitability of COTS devices
for on-board use in future space missions. 
In this section,
we report representative works
for each type of COTS processor.  

\subsection{Field-Programmable Gate Arrays (FPGAs)}
For high-performance solutions,
COTS FPGAs are preferred,
as they outperform both their space-grade counterparts \cite{access}
and the general-purpose CPUs \cite{aiaa}.
The authors of \cite{mpsoc}
propose a runtime reconfigurable SW/HW on-board processor 
for VBN,
which provides autonomous adaptation during the mission. 
The architecture relies on the RTEMS real-time operating system
to apply reconfiguration and fault mitigation. 
The implementation is performed on
Xilinx's Zynq UltraScale+ MPSoC.
The Zynq-7000 SoC FPGA is used in \cite{lentarisTVID} 
to implement compute-intensive CV algorithms for satellite pose tracking. 
This architecture is based on a HW/SW co-design methodology
that exploits the full potential of the Zynq structure. 
Iturbe \emph{et al.} \cite{iturbe} 
propose a HW/SW infrastructure 
for instrument data handling and processing,
which is prototyped on Zynq-7000.
They also equip their design with several hardening techniques for fault mitigation (e.g., watchdogs). 
The on-board space computer of \cite{csp},
which is also based on the Zynq FPGA,
integrates several functionalities towards improving in-flight reliability (e.g., memory scrubbing).

\subsection{Graphics Processing Units (GPUs)}
COTS embedded GPUs have also gained the interest of the space community.
The GPU platforms 
can provide sufficient performance
with less development effort than FPGAs. 
Kosmidis \emph{et al.} \cite{gpu}
examine the applicability of mobile GPUs in the space domain,
covering both the software and hardware perspectives. 
In particular,
they analyze the algorithms and workloads of space applications
to identify their suitability for GPUs,
and they also perform a benchmarking evaluation on GPUs. 
In \cite{gpu2},
the authors evaluate two graphics-based computing methodologies
(OpenGL 2.0 and Brook Auto)
for safety-critical systems. 
Their benchmark is an application modeling a 
VBN scenario in which the aircraft performs rendez-vouz
with an object. 
Moreover, 
the literature includes FPGA-GPU co-processing architectures.
The hybrid FPGA--GPU architecture of \cite{gpu_fpga}
employs Nvidia's
Tegra X2/X2i GPU as main accelerator.
The heterogeneous architecture of \cite{gpu_fpga_vpu} 
integrates  
a SoC FPGA for SpaceWire I/O transcoding,
the AMD SoC (CPU \& GPU) for acceleration,
and optionally, a VPU for AI deployment.
The authors of \cite{nano1} deploy neural networks for object detection
along with image compression techniques on Nvidia's Jetson Nano GPU. 
More recently,
in \cite{kosmidis_2022}, embedded GPU platforms are evaluated for accelerating 
data compression standards for space applications. 

\subsection{Vision Processing Units (VPUs)}
The Intel's Myriad family of VPUs  
is systematically being evaluated 
by the space community. 
In \cite{ai_space}, 
Furano \emph{et al.} report results from the radiation testing of Myriad 2,
which involves different functional tests   
and characterization of all SoC's memories. 
According to their experiments, 
Myriad 2 remains fully functional
after exposed to a total ionizing dose of 49 krad(Si).
The same VPU is used 
to accelerate CV algorithms for VBN applications \cite{myriad2},
and specifically, 
it provides a throughput of up to 5 FPS for pose estimation on 1-MegaPixel images within the power envelope of 1W.
Agarwal \emph{et al.} \cite{ubo1}
evaluate Myriad 2 
for implementing a star identification neural network. 
The results show that 
Myriad 2 provides sufficient performance,
while consuming around 1W and retaining an accuracy of 99.08\%.
Moreover, 
Myriad 2 is equipped on-board in the
$\mathrm{\Phi}$-{S}at-1 CubeSat mission of the
European Space Agency
as DNN demonstrator for EO 
\cite{fsat_myriad, myriad2022}. 
Myriad 2 is also placed in custom boards for use in space.
In particular,
Ubotica's CogniSAT CubeSat board \cite{ubo3} is an 
AI inference and CV engine that exposes the compute of Myriad 2 to the payload developer.
In the same context, 
Myriad 2 is integrated as the main accelerator 
in the HPCB platform \cite{hpcb1}, 
which is a payload data processor board 
for mixed-criticality space avionics.  
The results show that
Myriad 2 can provide sufficient AI acceleration
with limited power consumption,
while it sustains remarkable I/O throughput \cite{hpcb}. 
To increase the reliability of Myriad 2 as part of a mixed-criticality architecture,
the authors of \cite{ftmyriad_2022} propose fault-tolerant techniques tailored to this SoC. 
Myriad X, 
which is the newest VPU and 
the successor of Myriad 2, 
has received less attention.
In \cite{ubo2}, 
the authors report results for
deploying neural networks classifiers on
Myriad X.
The networks are 
trained on Mars imagery from the Reconnaissance Orbiter and Curiosity rover,
and the average inference time is 16-20ms with power lying around 1.8-1.9W.

\section{The Myriad Vision Processing Unit}
\label{vpu}

\begin{figure}[!t]
    \centering
    \includegraphics[width=\columnwidth]{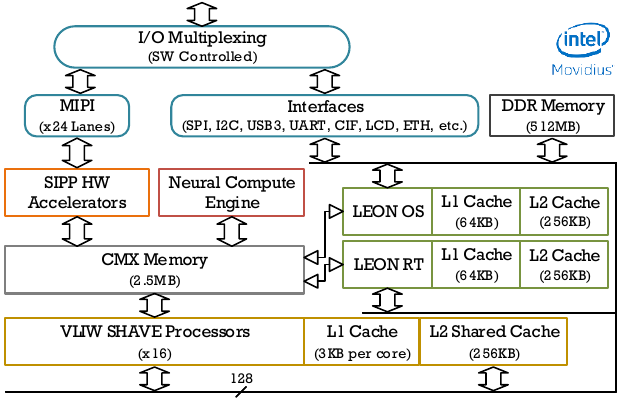}
    \caption{Intel's Myriad X VPU \cite{paralos}.} 
    \label{mx}
\end{figure} 

The Myriad family of VPUs includes multi-core COTS SoCs that 
target to 
terrestrial mobile and embedded applications \cite{myriad_vpu, myriad_hotc}.
The SoCs integrate
multiple peripherals for I/O,
hardware filters for low-power image processing,
and 
general-purpose \& vector processors.
All these components are 
connected to a multi-ported high-bandwidth shared memory hierarchy.
The heterogeneity of Myriad VPUs
allows for efficient mapping of AI and CV algorithms
with a power of 1-2W (depending on the VPU variant).
Besides their usage in space,
the Myriad VPUs are employed
for implementing 
DNNs \cite{myriad_dac2, myriad_tcad, movidius_ncs, movidius_ncs2},
machine learning algorithms 
(e.g., SVM classifiers \cite{marantos}),
and 
CV functions (e.g., stereo vision \cite{myriad_cv}).

In this work,
we employ the Myriad X VPU,
which is illustrated in Fig. \ref{mx}.
Contrary to Myriad 2,
this variant  
features a dedicated on-chip accelerator, 
called Neural Computer Engine (NCE),
for inferencing DNNs.
It also integrates
16 programmable 128-bit VLIW vector cores (called SHAVEs)
for accelerating CV functions,
hardware accelerators for imaging/vision tasks, 
and 2 general-purpose LEON4 processors
(called LEON OS and LEON RT).
Regarding the memory hierarchy, 
Myriad X provides 
on-chip 512MB LPDDR4 DRAM,
2.5MB scratchpad memory (called CMX),
as well caches for LEONs and SHAVEs.
Furthermore,
the SoC offers numerous imaging (MIPI lanes, CIF, LCD)
and general-purpose (e.g., SPI, USB, UART) I/O interfaces.

The Myriad X SoC is programmed via
the Myriad Development Kit (MDK), 
which includes all the necessary tools
(e.g., toolchain for the processors, C/C++ compiler, debugger)  
to implement custom CV and DNN algorithms. 
To deploy DNNs
of well-known frameworks (e.g., TensorFlow) 
on NCE,
the Intel's OpenVINO toolkit \cite{openvino} is used.
OpenVINO takes the frozen graph as input 
and employs tools such as the model optimizer and the Myriad compiler
to generate the programming file.
This binary network file is loaded on NCE via the mvNCI API,
which is also used to feed NCE with input data and receive its outputs in the form of tensors.


\section{AI/CV Acceleration for Lost-In-Space on Myriad X}
\label{prop}
\begin{figure}[!t]
    \centering
    \includegraphics[width=\columnwidth]{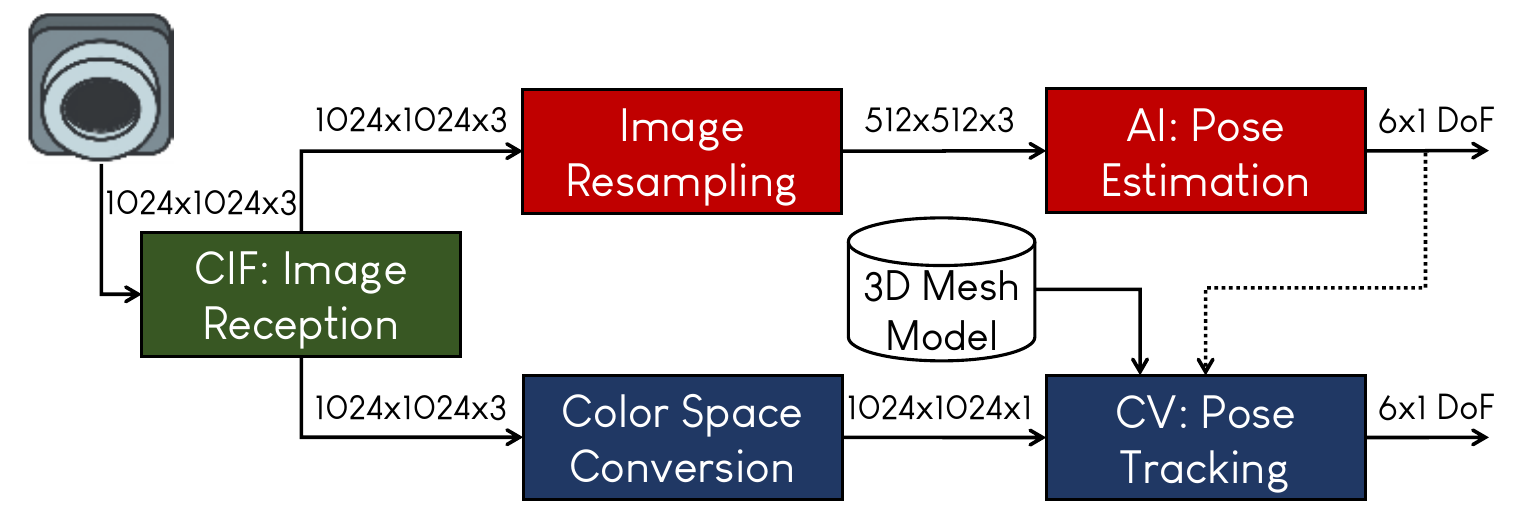}
    \caption{Cooperation of AI and CV for pose estimation/tracking on Myriad X.}
    \label{system}
\end{figure}

The targeted AI/CV system  
for pose estimation \& tracking
is illustrated in Fig. \ref{system}.
The input image is received by the Camera InterFace (CIF) of the VPU \cite{hpcb},
it is stored in DRAM,
and then it is subjected to pre-processing.
For the AI pipeline (upper block),
the pre-processing applies resampling
to reduce the image resolution, 
and thus 
facilitate DNN inferencing at the edge.
Our DNN is a mobile version of the UrsoNet network \cite{ursonet},
satisfying the real-time processing constraints.
For the CV pipeline (lower block),
the pre-processing converts the image to grayscale,
as required by our 5-stage CV algorithm \cite{lourakis, myriad2}.

The purpose of the current work is to 
quantify the costs and benefits of the two complementary algorithmic pipelines (CV and AI), 
fine-tune their implementation on Myriad X,
and
evaluate the potential gains of their acceleration on a single low-power SoC. 
We note that the final system integration
would also include
a high-level policy mechanism
to coordinate the two pipelines,
e.g., 
(i) decide when to provide an initial pose via AI
and when to continue tracking via CV,
or 
(ii) execute both pipelines for the same input
and decide which pose to keep; 
the creation of such a policy mechanism 
requires additional exploration with test campaigns
and customization to real satellite datasets
(e.g., to identify when the pose is lost).

\subsection{Pre-Processing}
To explore different trade-offs between DNN's performance and accuracy,
we employ three well-established resampling algorithms 
with increasing computational complexity.
First, Bilinear Interpolation
inputs 2$\times$2 pixel regions
with a stride equal to the reciprocal of the scaling factor and calculates 
the mean value per region. 
Second, Bicubic Interpolation
inputs 4$\times$4 pixel regions
and applies a cubic polynomial per region.  
Finally, 
Lanczos Interpolation
reconstructs the pixels through a Lanczos kernel (windowed sinc function), 
and then applies interpolation using convolution;
in this work, we consider 8$\times$8 pixel regions. 

The image resampling algorithms are accelerated on SHAVEs,
while LEON OS monitors the execution
and performs tasks such as
initializing the SHAVEs
and passing the entry points to them. 
At system level,
we configure the RTEMS Real-Time Operating System
(RTOS)
for LEON,
define the memory space of DRAM and CMX,
and set the clock of the SoC at the maximum frequency (700MHz).
In terms of high-level design,
the input image is divided into stripes,
i.e., segments of successive image rows,
and LEON OS assigns one stripe to each SHAVE 
to enable \emph{core parallelization}.
The input image is stored in DRAM (global memory),
and each SHAVE transfers its own stripe in a dedicated slice 
of CMX (scratchpad memory) via the SoC's DMA engine. 

Regarding low-level development details, 
the resampling algorithms are implemented in an iterative manner on SHAVEs.
In particular, each SHAVE performs a call to the selected 
resampling function per loop iteration, 
producing the output rows of its assigned image stripe one by one.
Therefore, the number of image rows
required to be transferred to each SHAVE's CMX slice per loop iteration 
is equal to the height of the 
algorithm's resampling region (e.g., 4 for Bicubic). 
Considering a scaling factor of 0.5 (i.e., stride equal to 2),
each loop iteration of Bicubic (4$\times$4 region) and Lanczos (8$\times$8 region)
needs to fetch only two new image rows.
To take advantage of this property, 
we employ a data structure
that implements 
a double-linked list,
where each node contains a row buffer,
and thus,
\emph{sliding buffer} is enabled 
by re-arranging the pointers that connect the nodes.
We also exploit the SoC's support 
for numerous data types
(e.g., ``u8'', ``u16'', and ``float16'')
to apply \emph{variable tuning}.
Moreover,
we enable \emph{SIMD computations}
by arranging the data into 128-bit vectors
and calling the corresponding routines 
to deploy additions and dot products on SoC's scalar arithmetic unit. 
The compiler is also enabled to 
map arithmetic operations onto the vector arithmetic unit.


\subsection{Deep Neural Network for Pose Estimation}
\begin{figure}[!t]
    \centering
    \includegraphics[width=\columnwidth]{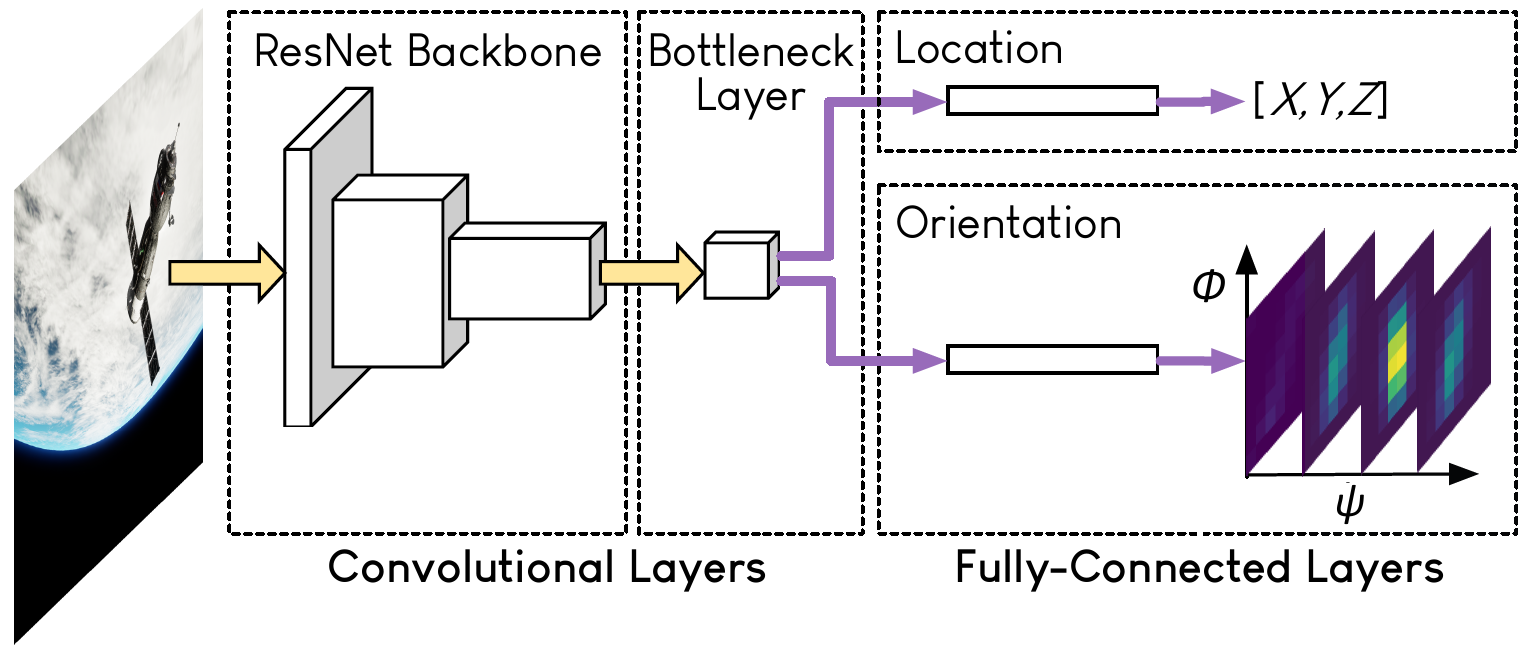}
    \caption{The architecture of the UrsoNet DNN for pose estimation \cite{ursonet}.}
    \label{dnn_archi}
\end{figure}

We employ the UrsoNet DNN \cite{ursonet} 
for estimating the satellite's pose.
The architecture of UrsoNet,
which is based on the ResNet-50 network, 
is illustrated in Fig. \ref{dnn_archi}. 
In comparison with the original ResNet architecture,
the global average pooling layer and last fully-connected layer
are removed.
In their place,
the designers of UrsoNet \cite{ursonet} insert
a bottleneck layer consisting of a 3$\times$3 convolution with stride 2
as well as two fully-connected layers calculating 
the satellite's location and orientation.
To build our network,
we use the ``soyuz\_hard'' dataset
and follow the training guidelines of \cite{ursonet}. 
The dataset is split using an 80-10-10 scheme, that is, 4000 images comprise the training set, while the validation and testing set consist of 500 of the remaining images each. 
The configuration of UrsoNet for deploying it on Myriad X
is reported in Table \ref{config}.
The initial weights of the backbone are acquired from a Mask R-CNN model pre-trained on the COCO dataset, as suggested in \cite{ursonet}. Data augmentation is performed by randomly rotating half of the training images per epoch. 
The image resolution is decreased
using the resampling algorithms and zero padding. 
All models are trained for 100 epoques on  
Nvidia's Tesla V100 GPU.

\begin{figure}[!t]
    \centering
    \includegraphics[width=\columnwidth]{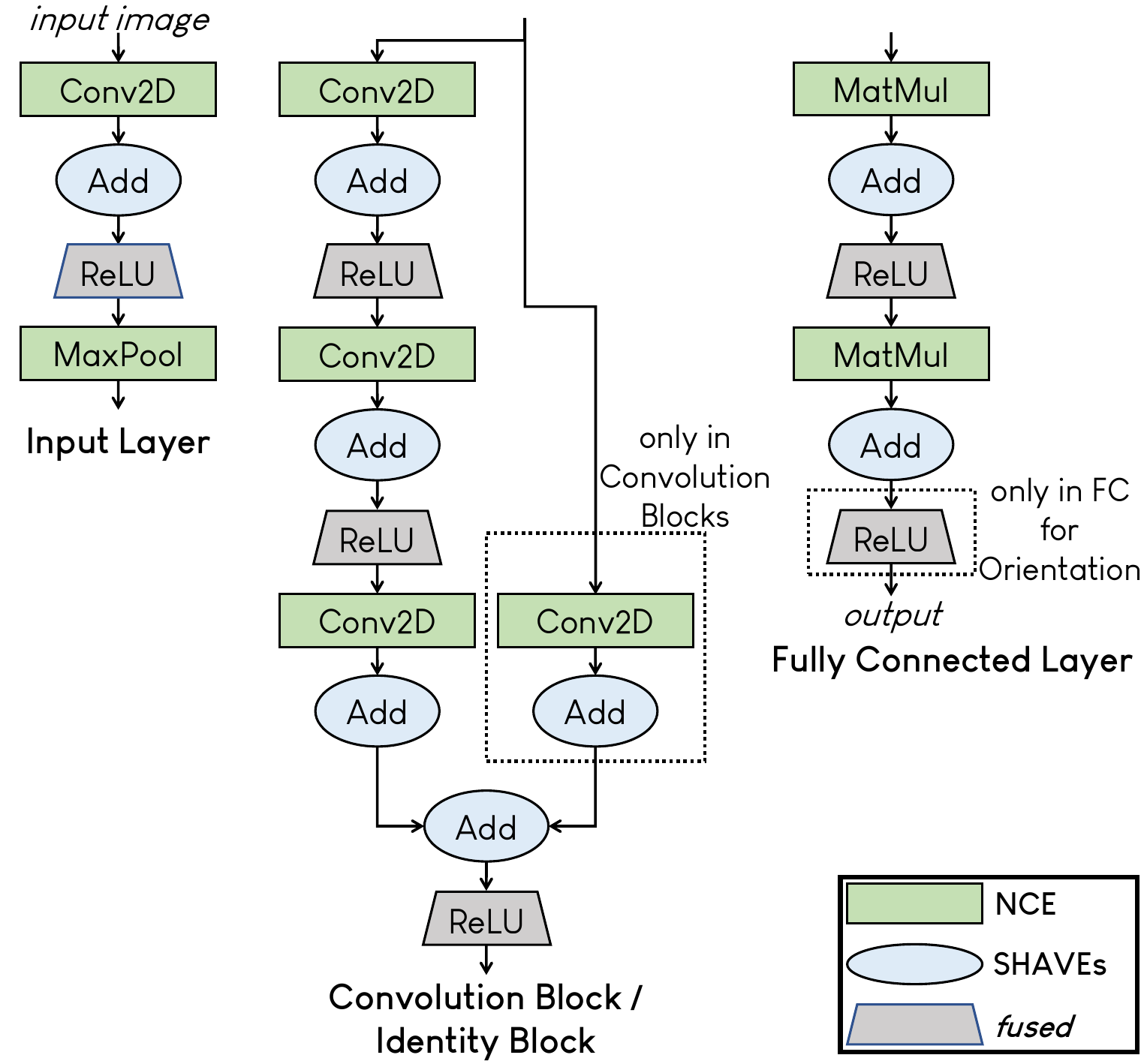}
    \caption{Mapping of the DNN for pose estimation on Myriad X.}
    \label{esti}
\end{figure}

To generate the binary network file and deploy it on the USB NCS2 
accelerator of Myriad X,
we follow the typical OpenVINO toolflow \cite{openvino}.
Fig. \ref{esti} illustrates the mapping of the basic network blocks of our UrsoNet architecture on Myriad X.
Compared to the original ResNet architecture, 
the batch normalization blocks are replaced by add operations,
which are accelerated on SHAVEs. 
Smaller operations, 
e.g., some permutations before the fully-connected layers,
are also executed on SHAVEs. 
The convolutions and fully-connected layers 
are mapped onto NCE,
while almost all the ReLU activation functions 
are optimized out by OpenVINO during the graph transformation stage,
i.e., they are fused with other operations. 

\begin{table}[!t]
\fontsize{9}{10}\selectfont
\renewcommand{\arraystretch}{1.15}
\setlength{\tabcolsep}{15pt}
\caption{The configuration of the UrsoNet DNN for deployment on Myriad X}
\label{config}  
\centering
\begin{tabular}{l c} 
\hline
 Backbone & ResNet-50 \\
 Bottleneck Width & $128$ \\
 Branch Size & $512$ \\
 Input Image & $1024$$\times$$1024$$\times$$3$ \\
 Resampling & Bilinear, Bicubic, Lanczos \\
 Inference Image & $512$$\times$$512$$\times$$3$ \\
 Ori./Loc. Resolution & $16$/$16$ \\
\hline
\hline
Pre-Trained Weights & COCO \\
 Dataset & ``soyuz\_hard'' \\
  Arithmetic & $16$-bit floating-point \\
 Epochs & $100$ \\
 Image Augmentation & Yes \\
 Optimizer & SGD \\
\hline
\end{tabular}
\end{table}

\subsection{Computer Vision Pipeline for Pose Tracking}
\begin{figure}[!t]
    \centering
    \includegraphics[width=\columnwidth]{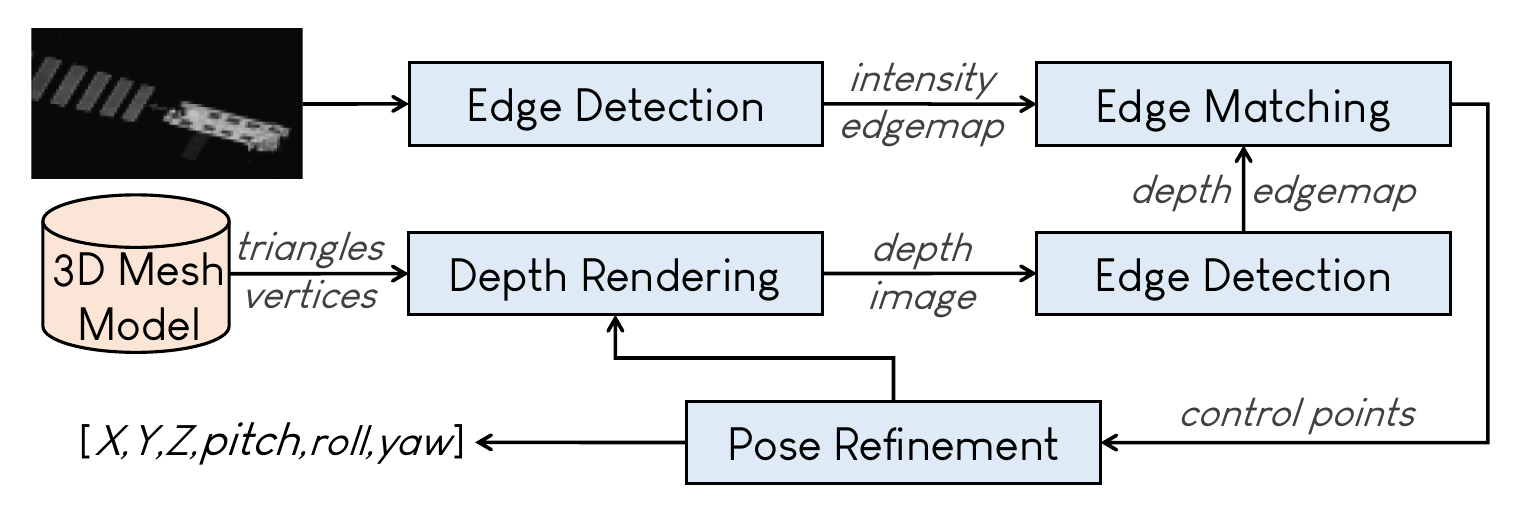}
    \caption{The block diagram of the CV pipeline for pose tracking \cite{lourakis, myriad2}.}
    \label{cvalgo}
\end{figure}

The continuous tracking of the satellite's pose is performed with a classic CV pipeline \cite{lourakis, myriad2},
which is illustrated in Fig. \ref{cvalgo}.
This algorithmic pipeline
involves functions for feature detection, perpendicular matching, image/depth rendering, and linear algebra operations.
In particular,
the pipeline evolves an initial pose
by matching edges
detected on the input image and a depth map (rendered from the satellite's 3D mesh model).
The matching edges 
are then used 
to refine the 6D pose
via robust regression calculations.
For the Edge Detection function,
we employ the well-established Canny edge detector \cite{canny},
which applies convolutions, non-maximum suppression and hysteresis thresholding.
The Depth Rendering function uses a triangle mesh model
and the 6D pose to generate the depth image,
whose pixels encode the distance 
between the camera and the nearest point on
the model’s surface.
The function is based on rasterization
to project the triangles on the image, 
and also involves tasks such as bounding box traversal and distance calculation.
Perpendicular Edge Matching
detects the correspondences between the input and the depth image,
and involves image traversal and comparisons.
Finally, 
Pose Refinement uses the matching edges and spatial information
to determine the change in satellite's pose
via linear algebra operations. 

\begin{figure}[!t]
    \centering
    \includegraphics[width=\columnwidth]{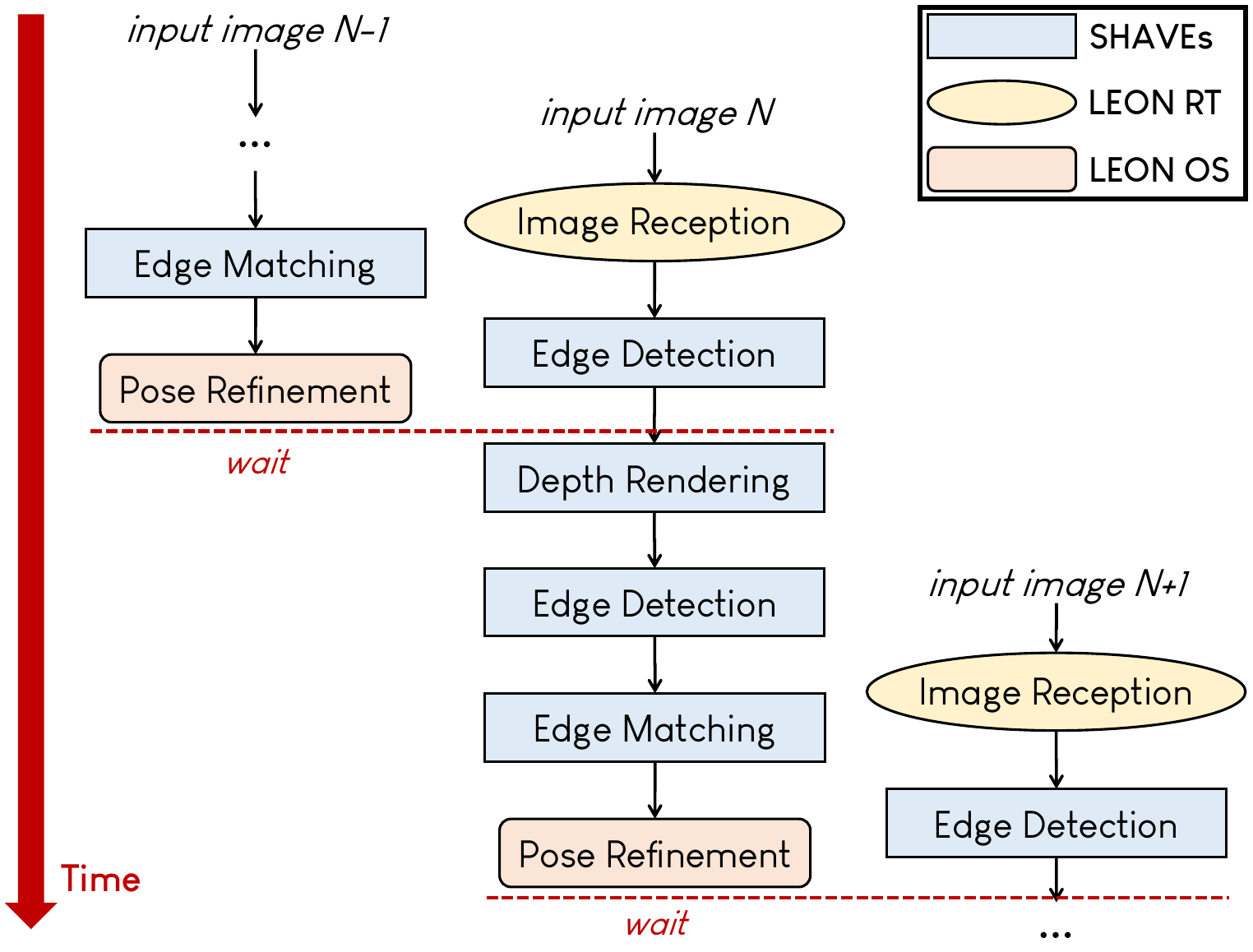}
    \caption{Scheduling of the CV pipeline for pose tracking on Myriad X.}
    \label{track}
\end{figure}

Considering that the CV pipeline 
is expected to run for multiple successive frames,
we integrate the function for image reception via CIF
in the pipeline. 
The heterogeneity of the SoC 
facilitates the complex scheduling 
presented in Fig. \ref{track},
where the reception of the next frame starts
before finishing the processing of the current frame.
The main functions, i.e., Canny Edge Detection, Depth Rendering, and Edge Matching,
are accelerated on SHAVEs, 
while LEON RT receives the image via the I/O interface
and LEON OS executes the algebraic-based Pose Refinement 
(not mapped onto SHAVEs due to limited BLAS/LAPACK library support). 


For the functions running on SHAVEs,
the processing is performed in image stripes (tiles of pixel rows).
Namely,
LEON OS divides the image into stripes
and assigns them to SHAVEs for parallel processing. 
We define the processing of one stripe as one task.
For Edge Detection and Edge Matching, 
we apply \emph{static task scheduling},
i.e., each SHAVE is assigned a predefined number of stripes to process.
In contrast,
to reduce the idle core time
in the content-dependent Depth Rendering,
we apply \emph{dynamic task scheduling},
i.e., each SHAVE is assigned a new stripe to process
upon finishing its previous one.
At lower level,
we apply the optimizations of the resampling algorithms,
i.e., variable tuning and SIMD,
as well as \emph{improved buffering} 
(to allow in-place computations), 
\emph{loop merging} 
(to perform calculations within the same loop), 
and \emph{cache optimization} 
(to improve data accesses).
All these techniques are manually developed in-house.


\section{Evaluation}
\label{exp}

\subsection{Experimental Setup} 
The development on Myriad X is performed
with MDK R15.4.
For the DNN implementation,
we use TensorFlow to train and build the model,
and OpenVINO 2021.3 to generate the network binary file. 
We consider 1-MegaPixel RGB image as input,
which is scaled to 512$\times$512 RGB for the DNN inference 
and transformed to 1-MegaPixel grayscale for the CV pipeline. 
We also employ a power management scheme
(using an API built on-top of the vendor routines)
to deactivate the SoC's idle components,
i.e., we power off the islands
of unused peripherals and hardware filters. 

\begin{table*}[!t]
\fontsize{9}{10}\selectfont
\renewcommand{\arraystretch}{1.15}
\setlength{\tabcolsep}{4.3pt}
\caption{Experimental results for the pose estimation AI pipeline on Myriad X 
(1024$\times$1024$\times$3 input image resampled to 512$\times$512$\times$3 for DNN inference)}
\centering
\begin{threeparttable}
\centering
\begin{tabular}{l  c c c  c c c c  c c  c c c c}
\hline
\multicolumn{8}{c}{\textbf{\textit{Image Resampling}}} &
\multicolumn{4}{c}{\textbf{\textit{DNN Inference}\tnote{4}}} & 
\multicolumn{2}{c}{\textbf{\textit{AI: Pose Estimation}}}\\ 
\cmidrule(lr){1-8} \cmidrule(lr){9-12} \cmidrule(lr){13-14}
\multirow{2}{*}{\textbf{Algorithm}} & \multicolumn{3}{c}{\textbf{Performance}} & \multicolumn{4}{c}{\textbf{Power}} & \multicolumn{2}{c}{\textbf{Performance}} & \multicolumn{2}{c}{\textbf{Accuracy}} & \multicolumn{2}{c}{\textbf{Performance}} \\
\cmidrule(lr){2-4} \cmidrule(lr){5-8} \cmidrule(lr){9-10}
\cmidrule(lr){11-12}\cmidrule(lr){13-14}
 & Latency & Improv.\tnote{1} & Speedup\tnote{2} & Core & DRAM & Total & Improv.\tnote{3} & Latency & Speedup\tnote{5} & LOCE\tnote{6}  & ORIE\tnote{6} 
 & Latency & Throughput \\
\hline
\hline 
Bilinear & 1ms  & 141$\times$ & 425$\times$ & 1.9W & 0.1W & 2.0W & 10\% & 373ms & 5$\times$ & 1.18m & 14.76$^\circ$ 
& 374ms & 2.7 FPS \\ 
Bicubic  & 6ms  & 68$\times$  & 285$\times$ & 1.9W & 0.1W & 2.0W & 9\%  & 373ms & 5$\times$ & 1.15m & 14.62$^\circ$ 
& 379ms & 2.6 FPS \\ 
Lanczos  & 19ms & 55$\times$  & 360$\times$ & 2.1W & 0.1W & 2.2W & 9\%  & 373ms & 5$\times$ & 1.15m & 14.57$^\circ$ 
& 392ms & 2.6 FPS \\ 
\hline
\end{tabular}
\begin{tablenotes}
\vspace{2pt}
\hspace*{-7pt}
\begin{minipage}[]{0.31\textwidth}
    \item[1] {\scriptsize improvement vs. straightforward SHAVE parallelization}
    \item[2] {\scriptsize speedup vs. execution on LEON4 CPU}
\end{minipage}\hspace{4.0pt}
\begin{minipage}[]{0.28\textwidth}
    \item[3] {\scriptsize improvement vs. default power management}
    \item[4] {\scriptsize synchronous inference (throughput = 1/latency)}
\end{minipage}\hspace{-6pt}
\begin{minipage}[]{0.4\textwidth}    
    \item[5] {\scriptsize speedup vs. original DNN (1024$\times$1024$\times$3 image)} 
    \item[6] 
    {\scriptsize LOCE=1.3m \& ORIE=10.7$^\circ$ in original DNN without resampling 
    \cite{ursonet}}  
\end{minipage}    
\end{tablenotes}
\end{threeparttable}
\label{ai}
\end{table*}

Regarding the AI pipeline,
we compare our final implementation of the resampling algorithms on SHAVEs
with the straightforward core parallelization scheme,
which 
divides the input image into 16 stripes
and the SHAVEs process one stripe in parallel 
without applying any of the proposed optimizations, 
and the execution on the LEON4 CPU
(columns 2--4 of Table \ref{ai}).
Furthermore,
we measure the power consumption 
and examine its reduction by our power management scheme (columns 5--8 of Table \ref{ai}).
Our resampled DNN inference on NCE \& SHAVEs
is evaluated by comparing it with
the respective execution on the same processors
for the original 1-MegaPixel image 
(columns 9--12 of Table \ref{ai}).
Finally, we examine the performance of the entire AI pipeline 
(columns 13--14 of Table \ref{ai}).
Regarding the CV pipeline, 
we evaluate its acceleration on the SHAVEs of Myriad X
in comparison with 
the respective optimized implementation on Myriad 2
and the execution on the LEON4 CPU of Myriad 2 
(columns 3--4 of Table \ref{cv}). 

\subsection{Experimental Results on Myriad X VPU}

Table \ref{ai} reports our results per AI pipeline and resampling algorithm. 
The focus is on performance (i.e., speed) and power.   
The latency of the pre-processing stage is 
1--20ms, depending on the algorithm,
which implies 55--141$\times$ performance improvement versus the straightforward parallelization on SHAVEs 
without the optimizations (sliding buffer, variable tuning, SIMD) and the use of the scratchpad memory (CMX).
Compared to LEON4, 
the speedup is remarkable,  
i.e., up to 425$\times$. 
In terms of power,
our manager improves the consumption by 10\%,
delivering around 2W (95\% of the total power is consumed by the LEON4 and SHAVE processors).
Moreover,
a single DNN inference (synchronous execution) 
on a 512$\times$512$\times$3 image requires around 373ms,
which is 5$\times$ faster than inferencing on the initial 
1-MegaPixel image.
At the same time, 
the mean Location Error (LOCE) is 1.15--1.18m 
and the mean Orientation Error (ORIE) is 14.57$^\circ$--14.76$^\circ$ for the
``soyuz\_hard'' dataset, 
i.e., both accuracy metrics are in the range of the respective values reported in \cite{ursonet}.
As a result, 
like in \cite{ursonet}, the accuracy of our DNN is close to that of the 
DNN without resampling (LOCE is 1.3m and ORIE is 10.7$^\circ$).
Furthermore, the network trained with the same  configuration on the ``soyuz\_easy'' dataset has a LOCE equal to 0.7m and an ORIE of 8.7$^\circ$. 
These values are sufficiently close to the results reported in \cite{ursonet}, namely 0.6m LOCE and 8.0$^\circ$ ORIE, 
however, once again the average latency is retained at 
373ms, indicating that training does not impact the performance.
In all cases, 
more fine-grained training
can provide better accuracy results for the same performance on the Myriad X SoC. 
Overall,
the entire AI pipeline achieves 2.6--2.7 FPS for synchronous execution (throughput will increase for asynchronous inferencing).

From the CV pipeline, 
Depth Rendering is the most demanding function,
providing a variable latency  
as it is highly dependent on the image content.
Our techniques
(dynamic scheduling, SIMD and cache optimization)
improve the latency of this function 
by 8$\times$ compared to the straightforward parallelization on SHAVEs.
Similar improvements are also delivered  
in the Edge Detection function. 
Edge Matching is a memory-intensive function involving comparisons,
thus, it is facilitated by our cache optimizations
to deliver a speedup of 4$\times$ compared to the straightforward SHAVE parallelization.
In Table \ref{cv}, we report the latency results 
for the CV pipeline along with the gains compared to the LEON4 and SHAVE processors of Myriad 2 \cite{myriad2}.
Compared to LEON4,
the speedup is 15--20$\times$
for the compute-intensive functions
and 
100$\times$ for the memory-intensive function.
Furthermore, 
compared to the SHAVE implementation on Myriad 2, 
we notice a speedup of 1.5$\times$ and 5$\times$, respectively, 
which mainly comes from the 4 extra SHAVEs
(16 in Myriad X)
and the increased clock frequency (700MHz in Myriad X).
Finally, in terms of accuracy,
the error is below 0.5m
and tracking is lost only in a few frames.

\begin{table}[!t]
\vspace{-5pt}
\fontsize{9}{10}\selectfont
\renewcommand{\arraystretch}{1.15}
\setlength{\tabcolsep}{6.2pt}
\caption{Performance of the pose tracking CV pipeline on Myriad X}
\centering
\begin{threeparttable}
\centering
\begin{tabular}{l | c c c}
\hline
\makecell[c]{\textbf{Function}}  & \textbf{Latency}  & \textbf{Speedup\tnote{1}} & \textbf{Speedup\tnote{2}} \\
\hline
\hline 
Intensity Edge Detection    & 24ms  & 15$\times$ & 1.5$\times$ \\ 
Depth Rendering             & 77--170ms & 20$\times$ & 1.5$\times$ \\
Depth Edge Detection        & 26ms  & 15$\times$ & 1.5$\times$ \\
Edge Matching               & 1ms  & 100$\times$ & 5$\times$ \\
Pose Refinement             & 90--120ms & 1.1$\times$ & 1.1$\times$ \\ 
\hline 
\textbf{\textit{CV: Pose Tracking}} & 218--341ms\tnote{3}  & 13$\times$ & 1.4$\times$ \\
\hline
\end{tabular}
\begin{tablenotes}
    \item[1] {\scriptsize speedup vs. LEON4 @600MHz (Myriad 2)}
    \item[2] {\scriptsize speedup vs. 12 SHAVEs @600MHz (Myriad 2)}
    \item[3] {\scriptsize sequential execution (without the scheduling of Fig. \ref{track})}
\end{tablenotes}
\end{threeparttable}
\label{cv}
\end{table}

\begin{table}[!t]
\fontsize{9}{10}\selectfont
\renewcommand{\arraystretch}{1.15}
\setlength{\tabcolsep}{2.9pt}
\caption{Performance of Lost-In-Space modules on Myriad X}
\centering
\begin{threeparttable}
\centering
\begin{tabular}{l | c c c}
\hline
\makecell[c]{\textbf{Module}}  & \textbf{Input Data} & \textbf{Latency} & \textbf{Throughput}   \\
\hline
\hline 
CIF: Image Reception\tnote{1}  & 1024$\times$1024$\times$3 & 63ms & 16 FPS \\ 
Pre-Processing            & 1024$\times$1024$\times$3 & 1--19ms & 50--1000 FPS \\ 
AI: Pose Estimation       & 512$\times$512$\times$3   & 373ms & 2.7 FPS \\
CV: Pose Tracking         & 1024$\times$1024$\times$1 & 218--341ms & 3.1--5.1 FPS\tnote{2} \\
\hline
\end{tabular}
\begin{tablenotes}
    \item[1] {\scriptsize CIF@50MHz}
    \item[2] {\scriptsize with function masking for streaming processing (scheduling, Fig. \ref{track})}
\end{tablenotes}
\end{threeparttable}
\label{sys}
\end{table}

\begin{table*}[!t]
\fontsize{9}{10}\selectfont
\renewcommand{\arraystretch}{1.15}
\setlength{\tabcolsep}{5.7pt}
\caption{Performance of embedded processors for the UrsoNet DNN (TensorFlow, inference on 512$\times$640$\times$3 image)}
\centering
\begin{threeparttable}
\centering
\begin{tabular}{l  c | c c c c }
\hline
\makecell[c]{\textbf{Device}}  & \textbf{Processor} & \textbf{Latency} & \textbf{Throughput} & \textbf{Power} & \textbf{Throughput-per-Watt}  \\
\hline\hline 
Intel Myriad X VPU (NCS2)                          & NCE + 16-core SHAVE @700MHz & 588ms   & 1.7 FPS & 5W\tnote{1}  & 0.34 FPS \\ 
\multirow{2}{*}{ARM Cortex-A57 CPU (Jetson Nano)}  & 4-core @1.4GHz              & 2830ms  & 0.4 FPS & 10W & 0.04 FPS \\ 
                                                   & 2-core @918MHz              & 7519ms  & 0.1 FPS  & 5W  & 0.02 FPS \\ 
\multirow{2}{*}{Nvidia Maxwell GPU (Jetson Nano)}  & 128-core @921MHz            & 761ms  & 1.3 FPS & 10W & 0.13 FPS \\ 
                                                   & 128-core @614MHz            & 958ms   & 1 FPS   & 5W  & 0.2 FPS \\  
\hline
\end{tabular}
\begin{tablenotes}  
    \item[1]{\scriptsize NCS2 (2W) hosted on Raspberry Pi 3 (3W).}  
\end{tablenotes}
\end{threeparttable}
\label{devices}
\end{table*}

Table \ref{sys} summarizes the performance of each module of the targeted AI/CV  embedded system. 
The image reception via CIF operating at 50MHz lasts 63ms 
according to our experiments with an FPGA transmitting data to the VPU
\cite{hpcb}.
The pre-processing time is negligible compared to the AI/CV execution,
which sustains a throughput of 2.7--5.1 FPS (sufficient for pose estimation on 1-MegaPixel images).
We note that further speedups can be achieved,
e.g., 
inferencing on 192$\times$256$\times$3 image delivers 15 FPS
with reasonable accuracy loss (LOCE is 1.9--2m),
while the throughput of the CV pipeline 
can be improved
by rendering a smaller image.

\subsection{Comparison to Embedded Devices}

In this section, 
we provide comparisons to other embedded devices
that are also considered for space avionics.
In particular, 
we compare our implementations on Myriad X with
the execution on 
CPUs (LEON, ARM),
GPUs (Jetson Nano, Tegra K1), and 
SoC FPGAs (Zynq). 

Starting with UrsoNet,
Table \ref{devices} presents the results
for inferencing on VPU, CPU, and GPU.
For this experiment,
we use input images that are scaled to
512$\times$640$\times$3 size.
The CPU is the ARM Cortex-A57
of Nvidia's Jetson Nano board,
while our GPU is the 128-core Maxwell of the same board.
For Jetson Nano,
we consider its two power modes:
the ``low-power'' at 5W
and the ``high-performance'' at 10W.
To make a fair comparison,
we consider the NCS2 VPU
hosted on a single board computer,
i.e., a Raspberry Pi 3,
and thus,
the total power consumption for inferencing is 5W.
In terms of latency,
the 4-core Cortex-A57 operating at 1.4GHZ is relatively slow,
i.e., NCS2 achieves a 6$\times$ speedup.
Compared to the GPU,
NCS2 achieves a slight improvement of 1.3$\times$,
but with 
half of the GPU's power consumption. 
When considering both
throughput and power,
NCS2 delivers
$\sim$1.7$\times$ more FPS per Watt than GPU
and
$\sim$8.5$\times$ more FPS per Watt than CPU.
In addition,
we provide comparison results for the original ResNet-50 network
(224$\times$224$\times$3 image),
which is the backbone of UrsoNet.
For this network, 
the Jetson Nano GPU  
provides increased throughput versus Pi 3 + NCS2,
i.e., 1.3--1.9$\times$ more FPS,
however,
when considering FPS-per-Watt,
the VPU inference outperforms the GPU
by 1.1--1.5$\times$.
Finally,
we note that if we use the Myriad X SoC 
rather than the pair Pi 3 + NCS2,
the proposed solution provides even better results against the competitor devices. 
In this case, the power consumption lies around 2--2.5W, 
while the latency does not include the USB communication time ($\sim$10ms).

Next,
we evaluate the acceleration of our CV functions
compared to implementations on other embedded devices.
Table \ref{biblio} summarizes the comparison with other works of the literature implementing the same functions.  
As already discussed 
in the previous subsection,
Myriad X outperforms its predecessor Myriad 2
by $\sim$1.5$\times$,
while compared to the general-purpose LEON4 CPU, 
it provides acceleration of 15--20$\times$
for the most demanding functions.
Compared to ARM Cortex-A9 of Xilinx's Zynq \cite{lentarisTVID},
we achieve remarkable speedup,
i.e., 
$\sim$14$\times$ for Canny Edge Detection (326ms on Cortex-A9),
$\sim$10$\times$ for Depth Rendering (869ms on Cortex-A9),
and
39$\times$ for Edge Matching (39m on Cortex-A9).
Compared to Nvidia's Tegra K1 GPU \cite{cannycomp},
Myriad X
provides 2$\times$ worse latency
for Canny Edge Detection
(12ms on Tegra K1 for 1-MegaPixel images).
However, 
in terms of performance-per-Watt,
Myriad X is 3$\times$ better
than Tegra K1 consuming 10W.

\begin{table}[!t]
\fontsize{9}{10}\selectfont
\renewcommand{\arraystretch}{1.15}
\setlength{\tabcolsep}{2.1pt}
\caption{Improvements by our CV implementations versus literature works}
\centering
\begin{tabular}{l c c c}
\hline
\makecell[c]{\textbf{Design}}  & \textbf{Device} &  \textbf{Improvement} & \textbf{Ref.} \\
\hline
\hline 
CV Functions     &   Myriad 2 & $\sim$1.5$\times$ speedup & \cite{myriad2} \\
CV Functions     & LEON4 & $\sim$15--20$\times$ speedup &  \cite{myriad2}\\
Edge Detection   & ARM Cortex-A9  & 14$\times$ speedup &  \cite{lentarisTVID}\\
Depth Rendering  & ARM Cortex-A9  & 10$\times$ speedup & \cite{lentarisTVID}\\
Edge Matching    & ARM Cortex-A9  & 39$\times$ speedup & \cite{lentarisTVID}\\
Edge Detection    & Tegra K1  & 3$\times$ performance-per-Watt & \cite{cannycomp}\\
CV Pipeline     &  Zynq-7000  & 3.5--4$\times$ better power & \cite{lentarisTVID}\\
CV Pipeline     & Zynq-7000 &1.4$\times$ performance-per-Watt & \cite{lentarisTVID}\\
\hline
\end{tabular}
\label{biblio}
\end{table}

Finally,
we compare our CV implementation
on Myriad X
to that on the programmable logic of Zynq-7000 \cite{lentarisTVID}.
The results show
that Myriad X trades off speed for power,
i.e.,
it is 2.5--3$\times$ slower than the FPGA
(considering the entire CV pipeline),
however,
it is 3.5--4$\times$ better
in terms of mean power consumption.
Furthermore,
it exhibits stable power consumption (2W) 
compared to the FPGA, which has a variation
between 2W and 9W.
Besides better power
and performance per Watt,
the Myriad X provides
additional benefits versus the FPGA.
First,
the large DDR memory of Myriad X
can facilitate in-flight programmability in space avionics,
i.e., store different algorithms/applications that can be seamlessly programmed.
On the other hand,
the FPGA requires the development of 
dynamic re-configuration techniques,
which also add timing penalties.
Second, to support AI acceleration 
and fast DNN deployment on
Zynq, 
the developer has to build Xilinx's Deep Learning Processor Unit (DPU). 
Therefore, 
considering the resources required for the DPU, 
a device like Zynq-7000 
is stressed
to implement the entire system 
(UrsoNet DNN + CV pipeline),
as the CV kernel already 
utilizes increased resources,
i.e., 36\% LUTs, 48\% DSPs, and 77\% RAMBs \cite{lentarisTVID}.

\section{Conclusion and Future Work}
\label{concl} 
The paper
showed that heterogeneous embedded SoCs
integrating diverse processors 
(general-purpose, vector, neural)
can provide promising AI/CV solutions
in challenging problems for space,
such as pose estimation in
lost-in-space and satellite tracking scenarios.
We combined low-level optimizations,
evaluated the capabilities of Myriad X,
and quantified the total costs of
each algorithm in the targeted AI/CV system. 
The results showed that the Myriad X
VPU
directly competes with other state-of-the-art processors,
widely considered for space applications,
offering promising throughput 
and significantly lower power consumption.
Our future work includes
improving the cooperation between 
the AI and CV pipelines by 
customizing on certain datasets/sequences
and fine-tuning the high-level SW mechanism 
to switch between pipelines in real time.

\section*{Acknowledgement}
The authors would like to thank
M. Lourakis \& X. Zabulis
for providing the initial software of the CV pipeline,
and
P. Proença \& Y. Gao
for providing the UrsoNet code in GitHub. 

\bibliography{REFERENCES}

\end{document}